\documentclass{article}
\usepackage{amsmath}
\usepackage{float,graphicx}
\usepackage{bm}

\usepackage[margin=1in]{geometry}
\usepackage{fancyhdr}

\usepackage{amsmath}
\usepackage{graphicx,psfrag,epsf}
\usepackage{enumerate}
\usepackage{natbib}
\usepackage{url}
\usepackage{subcaption}
\usepackage{amsfonts}
\usepackage{bm}

\usepackage{setspace}

\begin{document}

\title{\bf A Review and Comparison of Different Sensitivity Analysis Techniques in Practice}
\author{D Francom, A Nachtsheim\\
	Los Alamos National Laboratory
}
\maketitle

\section{Introduction}


There exist many methods for sensitivity analysis readily available to the practitioner. While each seeks to help the modeler answer the same general question -- \textit{How do sources of uncertainty or changes in the model inputs relate to uncertainty in the output?} -- different methods are associated with different assumptions, constraints, and required resources, leading to conclusions that may vary in interpretability and level of detail. Thus, it is crucial that the practitioner selects the desired sensitivity analysis method judiciously, making sure to match the selected approach to the specifics of their problem and to their desired objectives.

In this chapter, we provide a practical overview of a collection of widely used, widely available sensitivity analysis methods. We focus on \textit{global} sensitivity approaches, which seek to characterize how uncertainty in the model output may be allocated to sources of uncertainty in model inputs \textit{across the entire input space}. Generally, this will require the practitioner to specify a probability distribution over the input space. On the other hand, methods for \textit{local} sensitivity analysis do not require this specification but they have more limited utility, providing insight into sources of uncertainty associated only with a particular, specified location in the input space. (See \cite{turanyi2000local} and \citep{borgonovo2016sensitivity} for a detailed discussion of local sensitivity analysis methods.)

Our hope is that this chapter may serve as a decision-making tool for practitioners, helping to guide the selection of a sensitivity analysis approach that will best fit their needs. To support this goal, we have selected a suite of approaches to cover, which, while not exhaustive, we believe provides a flexible and robust sensitivity analysis toolkit. All methods included are widely used and available in standard software packages.

\subsubsection*{Selecting the Right Method}

\paragraph{Consider Available Resources}

We organize this chapter to adhere as closely as possible to a natural decision-making flow. In our experience, the first decision point often hinges on available computing resources, which dictate the number of model runs the modeler can allot to their sensitivity analysis and, therefore, the sensitivity analysis approach best suited: some sensitivity analysis methods require a large---or even an extreme---number of model runs, while others are designed to be informative and consistent with few model runs. 

To leverage approaches that require large numbers of runs, it is common for practitioners to rely on an emulator for the sensitivity analysis, rather than using the model of interest directly. Within this framework, the modeler need only execute the model runs necessary to build a sufficient emulator. The emulator---which requires vastly fewer computing resources---is then used for the sensitivity analysis. As we illustrate in the following section, for accurate emulators this approach tends to achieve comparable results to relying directly on the model of interest, while demanding far fewer resources. Of course, if the practitioner has immense resources at their disposal or if the model of interest is not expensive to run, they can instead carry out the sensitivity analysis on the model of interest directly if desired, and avoid any emulator inaccuracy from being propagated into the sensitivity analysis.

The sensitivity analysis method available to the practitioner is, therefore, dependent on whether the practitioner's resources allow at least enough model runs to build a sufficient emulator. Thus, we begin by partitioning the sensitivity analysis approaches considered into two groups: (1) those that require at least enough model runs to build an emulator and (2) those that may be used when the practitioner cannot obtain enough model runs to build a sufficient emulator.

\paragraph{A Note on Emulators}

Emulators provide a high degree of flexibility and can facilitate analyses including calibration and dimension reduction in addition to reducing the resources required for sensitivity analysis. The model runs required to build an emulator are selected via a space-filling design, but space-filling designs are not usually tied to particular types of emulators, so the emulator need not be decided upon at the time of design selection [SF design refs]. There are a variety of methods by which to construct emulators, such as:

\begin{itemize}
\item \textit{Gaussian process models (GPs)}, which are likely to perform better than alternatives when working with small (e.g., $10p$--$100p$, where $p$ is the number of inputs of interest) numbers of model runs. Variants exist that can scale to much larger sets of model runs. 
\item \textit{Bayesian additive regression trees (BART)} \citep{horiguchi2021assessing}, which have proven to be reliable models for moderate-sized data (e.g., $100p$--$1000p$).
\item \textit{Bayesian multivariate adaptive regression splines (BMARS)} \citep{francom2018sensitivity,francom2019inferring}, which also work well with moderate-sized data (e.g., $100p$--$1000p$) and have some advantages that make them especially useful for sensitivity analysis (via the R package BASS \citep{francom2020bass}),  as discussed in the following section. This is the emulator we use in all following emulator-based analyses.
\end{itemize}

\paragraph{Consider Constraints}

Once the practitioner has identified the resources available for sensitivity analysis, the next question to be answered is whether the inputs to the model can be varied independently. If not, the sensitivity analysis method selected must be able to accommodate dependent inputs. We consider methods for both independent and dependent inputs.

\paragraph{Additional Considerations}

After accommodating available resources and input dependencies, there will still likely be more than one appropriate sensitivity analysis methods available from which to choose. At this stage, we recommend that the practitioner consider what information is provided by each potential approach. Some features to keep in mind include:

\begin{itemize}
    \item Interpretability: Are method results intuitive to interpret? Does the method obtain summary quantities with meaningful units?
    \item Interaction Effects: Does the method provide insight into two-factor (or higher) interaction effects, when present?
    \item Effect Shapes: Does the method provide the ability to visualize effects? 
    \item Targeted Exploration: Does the method facilitate exploration of sensitivity when subsets of inputs are of interest, rather than sensitivity to individual input changes?
\end{itemize}

\noindent Many methods do not provide all of this information simultaneously. Rather, the practitioner must decide what properties are most valuable for the particular problem at hand and select a method accordingly or use multiple methods.

\subsubsection*{Illustrating the Methods}

We illustrate the practical use of a selection of sensitivity analysis methods using the material strength model introduced in \cite{preston2003model} and commonly referred to as the PTW strength model. We provide an introduction to the PTW strength model below.

\paragraph{PTW Strength Model}

Given a set of material-specific parameters, material-specific constants (e.g., atomic masses and other quantities with virtually no uncertainty), and a few experimental conditions, the PTW strength model produces a stress-strain curve, which is a key ingredient in many engineering and materials science models and analyses \citep{gray2005predicting}. The PTW strength model contains ten parameters, which are partly physically meaningful and partly treated as empirical tuning parameters. Thus, uncertainty about these parameters may come from both lack of physics understanding and lack of (or variation within) relevant data for tuning. Five of the PTW strength model parameters are also subject to a system of constraints. The experimental conditions required for the PTW calculation are: (1) the material's temperature; (2) the strain (change in length) applied to the material; and (3) the rate at which the strain occurs. The response of interest in the PTW strength model is stress (load on the material). Harder materials (and colder materials) will experience higher stress under a given strain. 

For the sensitivity analysis methods included, we use a simplified version of the PTW stength model. In the simplified version (1) we fix the experimental conditions at one combination of temperature, strain, and strain rate; and (2) we ensure that all values within parameter uncertainty ranges obey the parameter constraints. This allows us to perform all sensitivity analyses under the same settings so that results are comparable, though we will highlight when methods could be used without making these simplifications. Denote the simplified PTW strength model by $f(\bm x)$, where $\bm x$ is a $p$-vector, and $p=10$ for our analysis.

\paragraph{Chapter Overview}

In the next section of this chapter, we provide an overview of selected sensitivity analysis methods, focusing on practical considerations, and illustrate each via the PTW model. In the final section, we conclude with a brief discussion and summary of our recommendations for selecting sensitivity analysis techniques in practice.

\section{A Selection of Standard Approaches}

In this section we provide a practical overview of a selection of widely-used sensitivity analysis methods, following a natural decision-making process: (1) determine available resources; (2) determine input dependencies; (3) consider what information is provided by available methods. 

\subsection{At Least Enough Model Runs to Build An Emulator}

The following sensitivity analysis approaches should be used when the modeler has the resources to obtain at least enough model runs to build an accurate emulator. Methods to be used when inputs are independent are covered first; methods that accommodate input dependencies are taken up second.

\subsubsection*{Requires Independent Inputs}
\paragraph{Sobol' indices}

Sobol' indices is a variance-based method for sensitivity analysis that is widely-used because it provides both intuitive interpretation and flexibility. In addition to generally ranking the importance of inputs, with Sobol' indices, the modeler can (1) capture information about interaction effects (second-order and higher); (2) visualize individual effects to understand nonlinearities; and (3) calculate results on subsets of inputs for targeted exploration, all via a metric that is based on variance, which is widely understood and therefore easy to interpret. Note, however, that the method relies solely on variance to summarize uncertainty, which can be seen as a potential disadvantage, since higher order moments may provide added information. In addition to requiring independence in the inputs (or subsets of inputs), this method also can require vast numbers of model runs, making it absolutely necessary that the modeler selects this method for use only when they have resources for many model runs or for at least enough model runs to build an accurate emulator.

\subparagraph{Mathematical Details}

Sobol' indices \citep{sobol2001global} rely on the functional ANOVA decomposition, which decomposes the function $f(\bm x)$ into a sum of the overall mean, main effects, two way interactions, three way interactions, etc.,
\begin{align}
\label{eq:fanova}
    f(\bm x) = f_0 + \sum_{i=1}^p f_i(x_i) + \sum_{i=1}^{p-1}\sum_{j>i} f_{ij}(x_i,x_j) + \dots + f_{1\dots p}(\bm x).
\end{align}
Inputs are assumed to be independent random variables so that $\bm x \sim \pi(\bm x)$ where $\pi(\bm x) = \pi_1(x_1)\dots \pi_p(x_p)$. The terms in the decomposition, $f_u(\bm x_u)$ for $u \subseteq \{1,\dots,p\}$ are constructed to be orthogonal and zero-centered with recursive definition 
\begin{align}
    f_u(\bm x_u) = E\left[ f(\bm x) | \bm x_u \right] - \sum_{v \subset u} f_v(\bm x_v)
\end{align}
where, when $u$ is the empty set, we define $f_u(\bm x_u) = f_0 = E\left[ f(\bm x) \right]$. For example, consider $u=\{ 1 \}$.  If $f$ is not purely additive in this variable, the effect of variable 1 will be different depending on the values of the other variables.  Then $f_1(x_1)$ can be thought of intuitively as the average effect of variable 1, where the average is across variables $2,\dots,p$, with the overall average $f_0$ removed.  If $u=\{ 1,2 \}$, the two way interaction $f_{12}(x_1,x_2)$ is the average effect of variables 1 and 2 together, with the main effects for variables 1 and 2 removed and the overall mean removed, meaning that this is the interaction after the main effects have been accounted for.  The function decomposition in Equation \ref{eq:fanova} becomes a variance decomposition in four steps: (1) square both sides, (2) subtract the $f_0^2$ term from both sides, (3) take the expectation of both sides (the orthogonality of the effects is important here, as it makes the expectation of most terms zero), and (4) use the identity $Var(z) = E(z^2) - E(z)^2$ (the effects having zero mean is important here), resulting in
\begin{align}
    Var\left[f(\bm x)\right] = \sum_{i=1}^p Var\left[f_i(x_i)\right] + \sum_{i=1}^{p-1}\sum_{j>i} Var\left[f_{ij}(x_i,x_j)\right] + \dots + Var\left[f_{1\dots p}(\bm x)\right].
\end{align}
To obtain Sobol' indices, we then divide both sides by $Var\left[f(\bm x)\right]$ and interpret sensitivity as the proportion of response variance explained by an input or interaction.

\subparagraph{Total indices: A single number summary}

Another important use of Sobol' indices is in the formation of \textit{total indices} for each input, which serves as a single number summary of a variable's contribution. Total indices are obtained by adding a variable's main effect and all interactions that include it, resulting in a metric that accounts for the main effect and interactions of all orders.  Total indices are relatively intuitive to interpret: when not divided by total variance, the total effect for variable $i$ can be interpreted as the expected remaining variance in the response \textit{when all inputs other than $i$ are known}. However, total indices no longer quantify \textit{individual} interactions. Still, using total indices, a modeler can gain insight into relative importance of variables and the presence of interactions: variables with larger total effects are relatively more important and variables for which the total effect is greater than the main effect are involved in interactions. 

Extensive research has been conducted in the area of calculation of Sobol' indices. Due to the computational burden, most research has focused on estimating main effect sensitivity and total effects. The Fourier Amplitude Sensitivity Test (FAST) \citep{cukier1973study,saltelli1999quantitative}, which can be used to obtain Sobol' indices, seeks to reduce the computational burden by transforming into Fourier space to reduce the dimensionality of the integrals. FAST is often quite accurate, making it a good choice to reduce required model runs.  Because low dimensional integrals are easier to approximate, extensive work has been done to reduce many of the Sobol' integrals to be low dimensional for GPs \citep{oakley2004probabilistic,marrel2009calculations}. For some other emulators (e.g., polynomial chaos \citep{sudret2008global}, Bayesian additive regression trees (BART) \citep{horiguchi2021assessing}, Bayesian multivariate adaptive regression splines (BMARS) \citep{francom2018sensitivity,francom2019inferring}, and various other tensor-product basis models \citep{chen2005analytical}), the integrals leading to Sobol' indices are analytical under particular input distribution choices. 
Generalizations of the Sobol' decomposition that allow for input dependence have also been explored (see, for instance, the discussion in \cite{iooss2019shapley}).

\subparagraph{Application to the PTW Strength Model}

\begin{figure}[htbp]
         \centering
         \includegraphics[width=\textwidth]{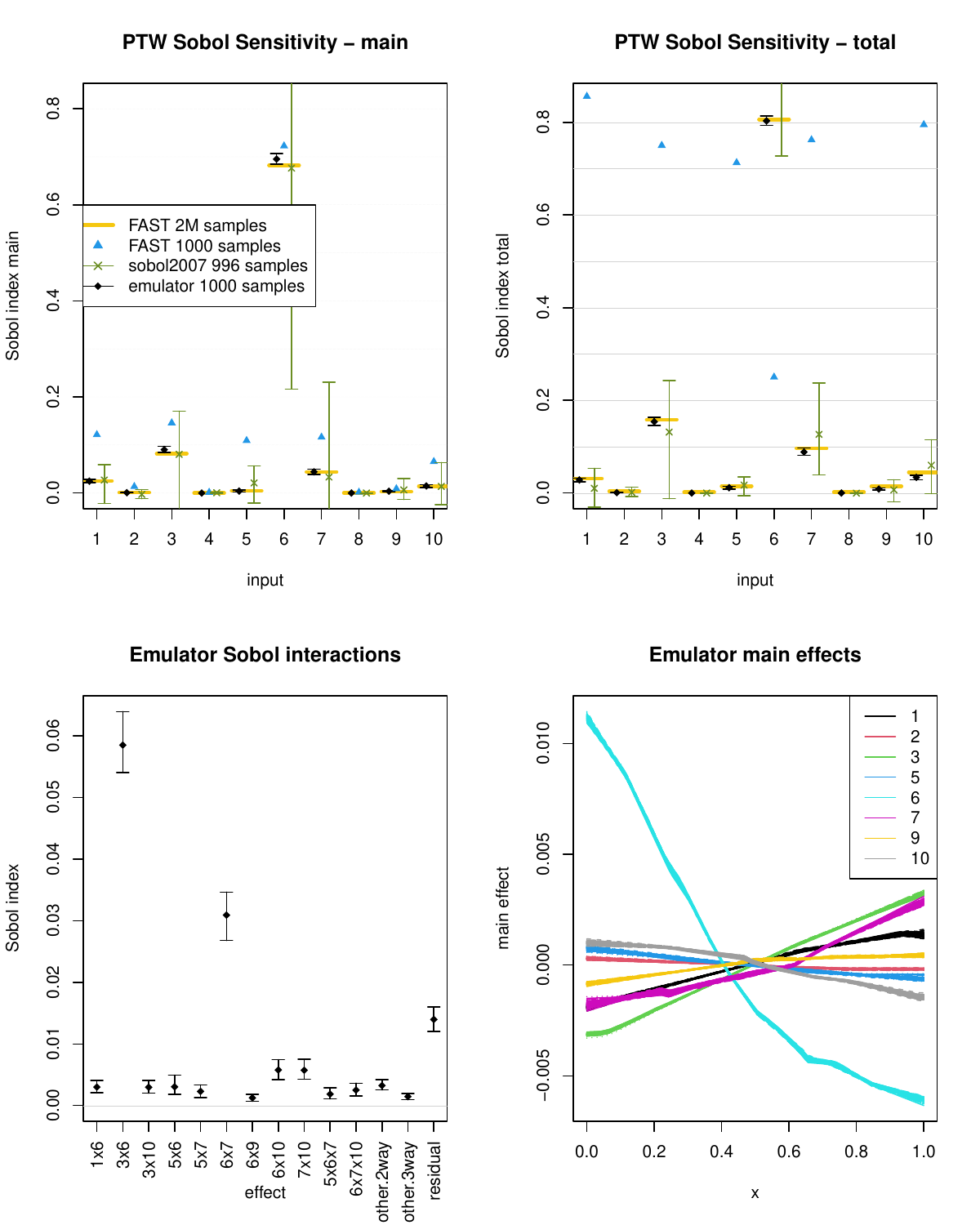}
         \caption{Sobol' sensitivity analysis for PTW. The top two plots show the main and total effects comparing a ``ground truth'' (FAST method with two million model runs) to FAST, a pick-freeze algorithm (\texttt{sobol2007} from R package \texttt{sensitivity}), and emulator-based results (\texttt{sobol} from R package \texttt{BASS}) all using about 1000 model runs. FAST with limited model runs is unreliable, while sobol2007 is fairly reliable but with large uncertainty. The emulator-based approach is accurate at 1000 model runs. The bottom plots show the Sobol' indices for interactions (as well as emulator residual, or unexplained variance) and main effects using the emulator-based sensitivity analysis.}
         \label{fig:sobol}
\end{figure}

We illustrate the use of Sobol' indices for the PTW strength model in Figure \ref{fig:sobol}. The four plots show the Sobol' indices and effects using an emulator with 1000 PTW model evaluations (via the \texttt{sobol} function in the \texttt{BASS} R package). Uncertainty for the emulator-based approach comes from doing the Sobol' decomposition for each emulator posterior sample. The top two plots also show Sobol' indices obtained without an emulator using the \texttt{fast99} and \texttt{sobol2007} functions from the R package \texttt{sensitivity}. For main and total indices, the emulator approach with 1000 model runs matches the FAST approach with two million model runs. The FAST approach with 1000 model runs is unreliable (especially for total indices), while the \texttt{sobol2007} (which relies on so called pick-freeze sampling tricks to estimate the expectations of the Sobol' indices with fewer function evaluations) approach with nearly 1000 model runs has relatively large uncertainty. 

The left plot of Figure \ref{fig:sobol} indicates that about 70\% of the variation in PTW response is a result of variation in input 6 on its own. Main effects for inputs 3 and 7 explain about 10\% and 5\% of the variation in PTW response, respectively. Variation in inputs 4 and 8 explain no variance. The bottom left plot shows proportion variation in the PTW response due to different two and three way interactions, as well as unexplained (residual) variation (obtained in comparable form by taking the emulator estimated residual variance divided by the total variance of the emulator mean, $Var\left[f(\bm x)\right]$). These indicate that there is an interaction between inputs 3 and 6 that explains 6\% of the variation in PTW response after accounting for the main effects for inputs 3 and 6. There is an interaction between inputs 6 and 7 that explains 3\% of the output variation, and all other effects are small relative to the unexplained variation. The bottom right plot shows the main effects $f_i(x_i)$ over the various ranges of the variables (so that each variable $x_i$ is scaled to be between 0 and 1). This shows that increasing input 6 decreases the PTW response in a nonlinear way. The magnitude of the input 6 main effect can be seen to be much larger than the magnitude of the other main effects, which aligns with the top left plot.

\begin{figure}[htbp]
         \centering
         \includegraphics[width=\textwidth]{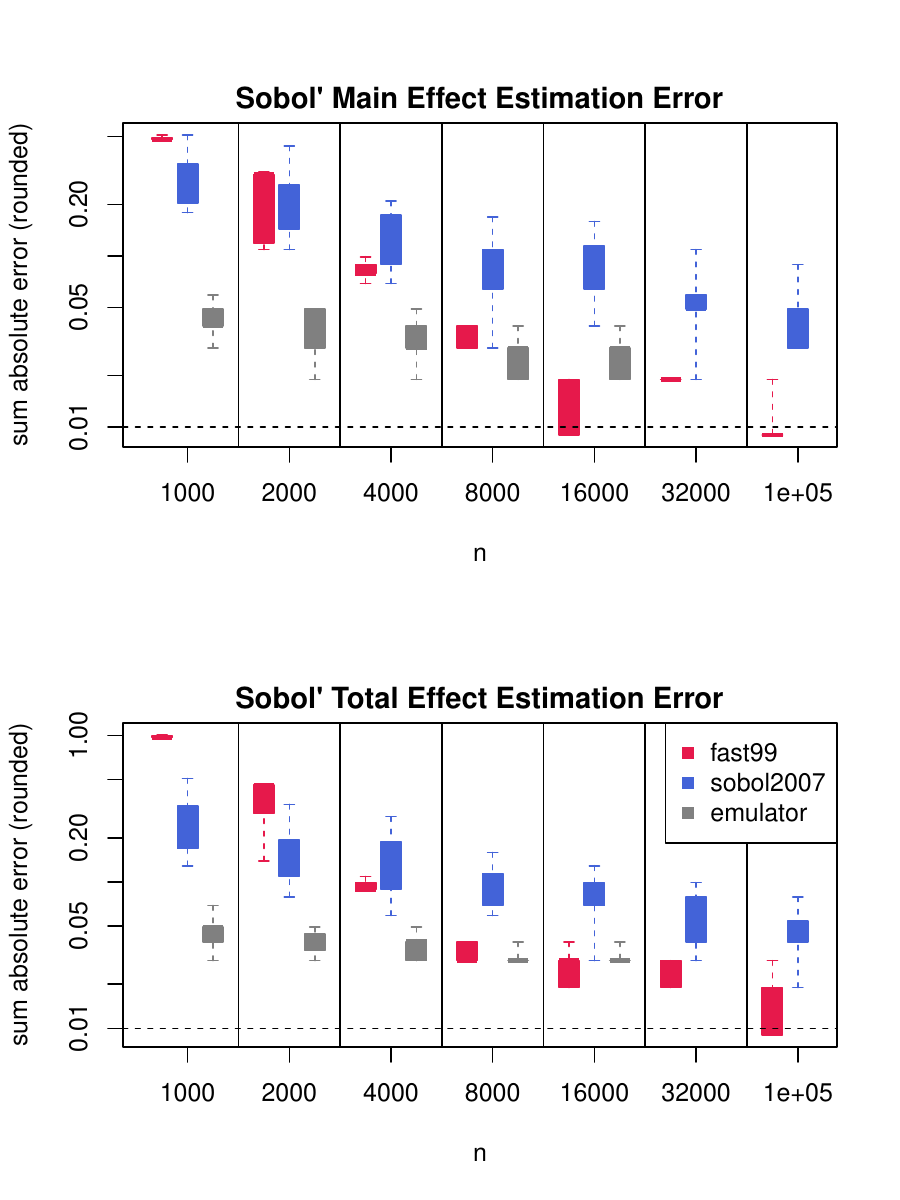}
         \caption{PTW Sobol' index estimation error as a function of number of model runs. Boxplots show results from 20 replications under each sample size. The emulator-based approach is far more accurate with limited model runs, with the FAST method catching up only with more than ten thousand model runs. As model runs increase, the FAST method is more accurate than pick-freeze style algorithms like the \texttt{sobol2007} function from R package \texttt{sensitivity}.}
         \label{fig:sobol_converge}
\end{figure}

The FAST approach with two million model runs is almost certainly accurate at estimating Sobol' indices when rounded to two decimal places. In Figure \ref{fig:sobol_converge}, we show how accuracy of the \texttt{fast99}, \texttt{sobol2007}, and emulator approaches converges as the number of model evaluations increases.  We round the estimated Sobol indices to two decimal places and sum their absolute error when compared to the FAST approach with two million model runs, also rounded to two decimal places. We repeat this twenty times with different random seeds for each sample size. Total absolute error of 0.01, represented with a dotted line, indicates that only one of the Sobol' indices was wrong and only by 0.01. As sample size approaches one hundred thousand, the FAST approach achieves total absolute error of zero (to two decimal places), which we represent as slightly less than 0.01 on the plot in order to not extend the log-scale plot axis unnecessarily. From \ref{fig:sobol_converge} we see that, for the PTW strength model, the emulator option is most accurate (for both main and total effects) until we can obtain tens of thousands of model runs, after which it is overtaken by FAST.

\subsubsection*{Allows Dependent Inputs}
\paragraph{Shapley Values}

Shapley values were developed in a game theory context with the intent to quantify how much value a player brings to a cooperative game with other players.  The analogy to sensitivity analysis comes when we replace players in a game with input variables in a model \citep{owen2014sobol,song2016shapley}. 

Shapley values can take a variance-based approach and so retain some of the interpretability of the Sobol' decomposition. Extensions to the method allow the modeler to explore the presence of interactions \citep{sundararajan2020shapley,rabitti2019shapley} and sensitivity for subsets of inputs. Unlike Sobol' indices, Shapley values allow for input dependence, handling this case in such a way that the sum of the effects still equals the total variance. However, like Sobol' indices, this method requires a large number of model runs or the use of an emulator. Further, even when the model runs are complete, the Shapley values can still be expensive to compute in some software.

\subparagraph{Mathematical Details}

The Shapley value is defined as 
\begin{align}
    Sh_i = \frac{1}{p}\sum_{u\subseteq \{1,\dots,p\}\backslash \{i\}} \begin{pmatrix}
  p-1 \\ 
  |u| 
\end{pmatrix}^{-1} \left[ val(u \cup {i}) - val(u) \right],
\end{align}
implying that for every subset of variables $u$ that excludes variable $i$, we find the value that variable $i$ adds if it were included in the given subset, $val(u \cup {i}) - val(u)$. The measures of added value for variable $i$ are taken in a weighted sum, where the weights account for the differing numbers of subsets of a particular size. 

The choice of the value function $val(\cdot)$ varies, but \cite{owen2014sobol} chooses $val(u) = Var\left[E\left[f(\bm x) | \bm x_u \right] \right]$ which is a function of the terms in the Sobol' decomposition discussed previously, $Var\left[E\left[f(\bm x) | \bm x_u \right] \right] = \sum_{v\subseteq u} Var\left[ f_v(\bm x_v) \right]$.  These Shapley values are non-negative, are zero for variables that never add value, and have other useful properties described in \cite{song2016shapley}. When inputs are independent, they are bounded by Sobol main and total indices. The main advantage of Shapley values as a single number summary is that they account for interactions by distributing them equitably among input variables involved in the interaction. Hence, as opposed to the Sobol' total indices which count interactions multiple times and so lose interpretability, Shapley values maintain the property that they sum to the total variance $Var\left[f(\bm x)\right]$. Another choice of value function is $val(u) = Var\left[E\left[f(\bm x) | \bm x_u \right] \right]/Var\left[f(\bm x)\right]$, in which case the Shapley values sum to one rather than the total variance \citep{song2016shapley}. In the case of dependent inputs, $Var\left[E\left[f(\bm x) | \bm x_u \right] \right]$ can still be calculated. The desirable properties of variance decomposition no longer hold for the Sobol' index, but Shapley values are still meaningful as they equitably distribute effects of correlated inputs to the individual Shapley values \citep{owen2017shapley,iooss2019shapley}.

Variations on the Shapley values defined above can be used for quantifying the importance of particular interactions \citep{rabitti2019shapley}.  A different choice of value function $val(u, \bm x_{u}) = E\left[f(\bm x) | \bm x_u \right]$ leads to the approach of \cite{lundberg2017unified} (SHAP), which presents the Shapley values as a function of $\bm x$, allowing the values to be visualized so the modeler can gain better understanding of the function. This approach is popular in machine learning applications.

\subparagraph{Application to the PTW Strength Model}

\begin{figure}[htbp]
         \centering
         \includegraphics[width=\textwidth]{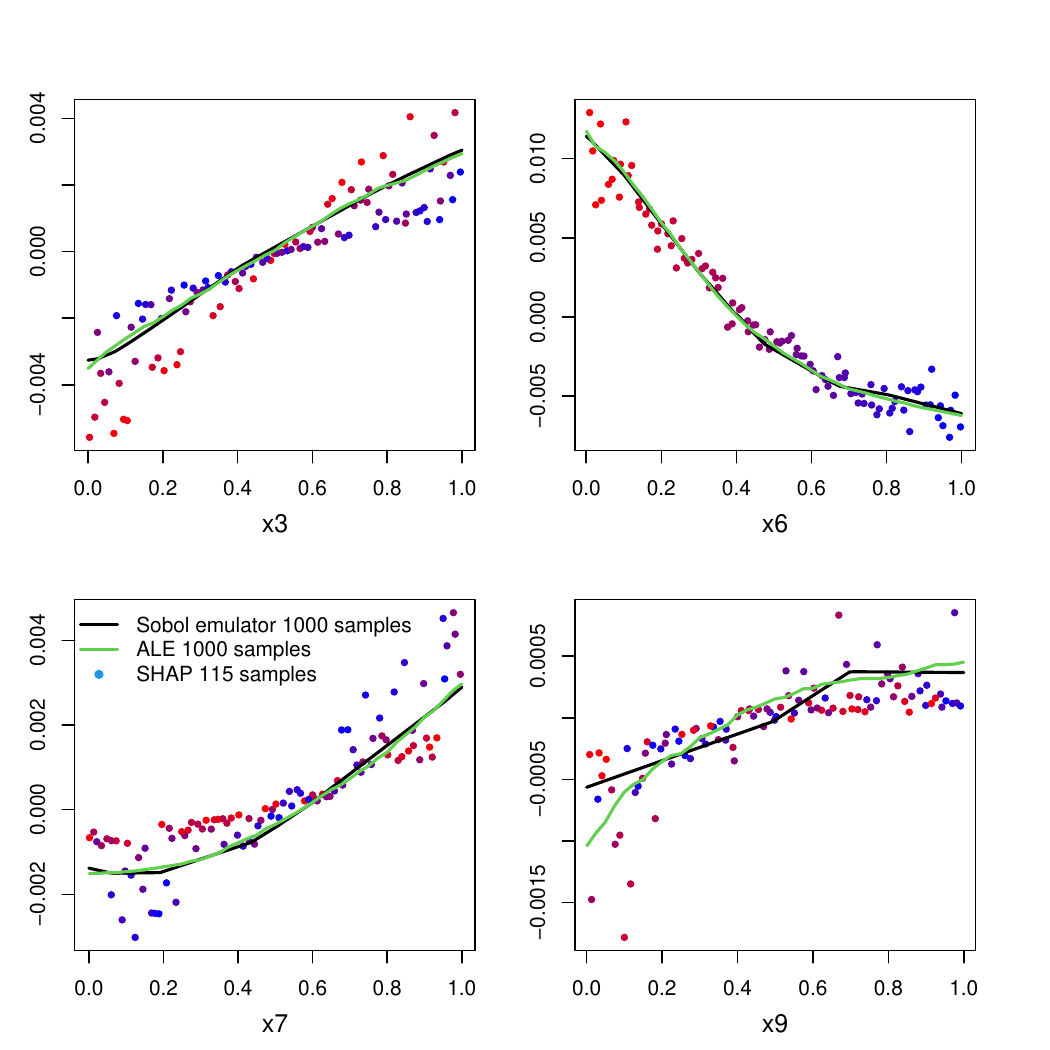}
         \caption{Effects visualized with three different methods. The Sobol' effects are $f_i(x_i) = E[f(\bm x) | x_i]$ using an emulator. The Shapley values are using $val(i, x_{i}) = E[f(\bm x) | x_i]$ to get $Sh_i(x_i)$ estimated from model evaluations directly. The points are colored according to $x_6$, with which $x_3$ and $x_7$ clearly interact. Accumulated local effects (ALE) use the PTW model evaluations directly. Sobol' and ALE two-way interaction effects can also be visualized, but that is omitted here.}
         \label{fig:effects}
\end{figure}

\begin{figure}[htbp]
         \centering
         \includegraphics[width=\textwidth]{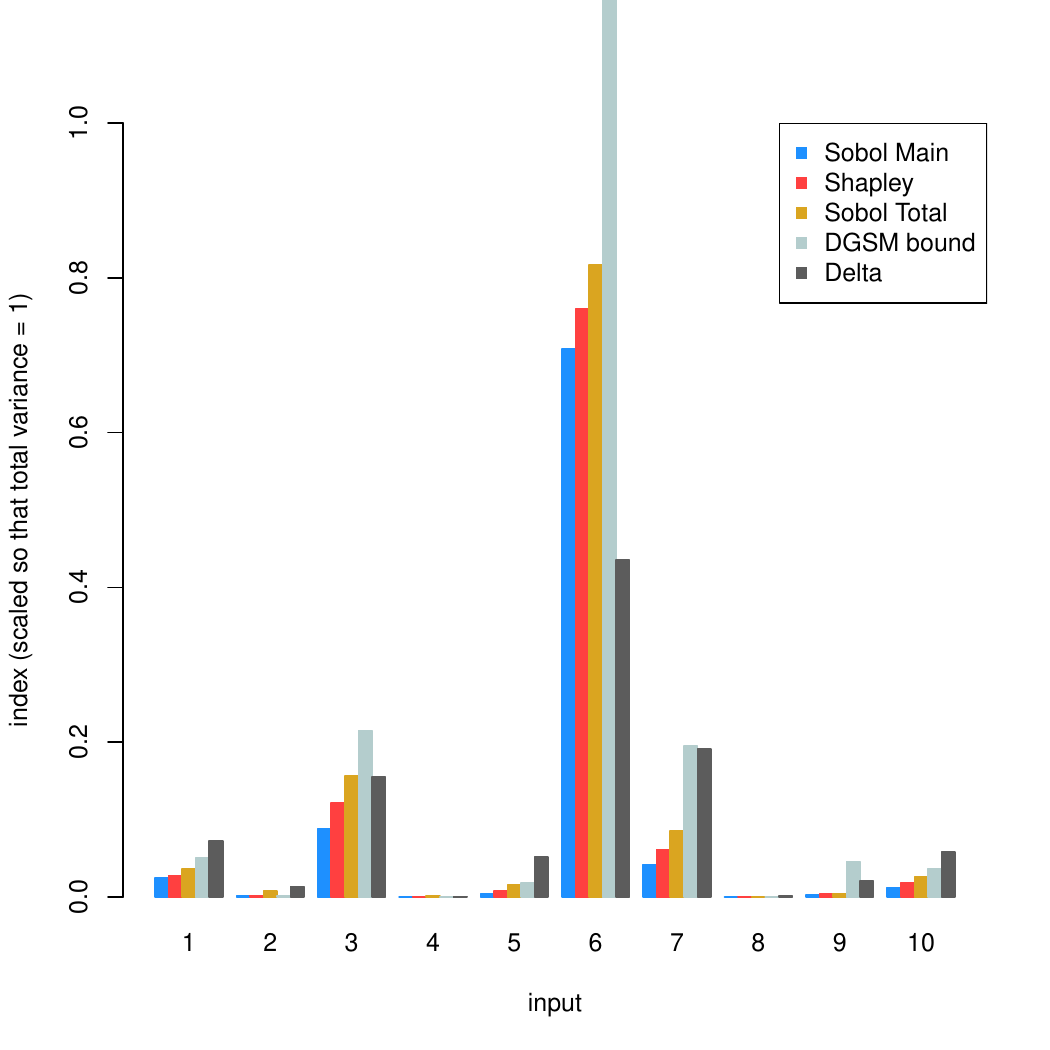}
         \caption{Comparison of different sensitivity metrics for PTW. The $y$-axis is scaled such that the total PTW variance is 1 (i.e., $Var\left[f(\bm x)\right]=1$) so that Sobol, Shapley, DGSM, and Delta sensitivity indices are comparable. Sobol total indices are larger than first order indices (or main effect indices) indicating interactions, and Shapley values fall between the Sobol first order and total indices. DGSM indices provide an upper bound for the Sobol total indices. Delta indices reveal different sensitivities because of the different way in which they are constructed--they find sensitivity to changes in output uncertainty beyond variance. An emulator was used to obtain the Sobol-based indices and delta indices in this plot, while the DGSM was obtained using 200 model evaluations.}
         \label{fig:shap_sob_del_dgsm}
\end{figure}

In illustrating Shapley values for the PTW strength model, we employ the \texttt{shapleySubsetMc} function from R package \texttt{sensitivity} which uses  $val(u) = Var\left[E\left[f(\bm x) | \bm x_u \right] \right]/Var\left[f(\bm x)\right]$. Figure \ref{fig:shap_sob_del_dgsm} shows results in comparison with other single number summaries. We also employ the python library \texttt{SHAP} which uses $val(u, \bm x_{u}) = E\left[f(\bm x) | \bm x_u \right]$ to visualize effects. Figure \ref{fig:effects} shows these results in comparison to other effect visualization methods.

\paragraph{Delta Indices}

The delta sensitivity index \citep{borgonovo2007new} is a powerful and flexible metric that moves beyond the variance-based approaches taken by Sobol' indices and Shapley values. Rather than relying solely on variance to summarize uncertainty, the delta index approach considers the relationship between changes in the inputs and changes in the \textit{entire probability density} of the response to measure sensitivity. 

Though expected change in density is not as interpretable as the Sobol' decomposition, this is a powerful metric because it uses the entire density of the response.  Computationally, this metric requires many model runs to compute, but work has been done to enable estimation of the metric with a given (i.e., not specially designed) set of model runs \citep{plischke2013global}.  This metric does not provide shapes of particular main effects, but instead can provide variations of response densities.  While it does not quantify particular interactions, it uses all interactions when determining the metric.  This metric can be used for cases with dependent inputs, subsets, and monotone transformations of the response, making it one of the more flexible metrics that we consider. 

\subparagraph{Mathematical Details}

\begin{figure}[htbp]
         \centering
         \includegraphics[width=\textwidth]{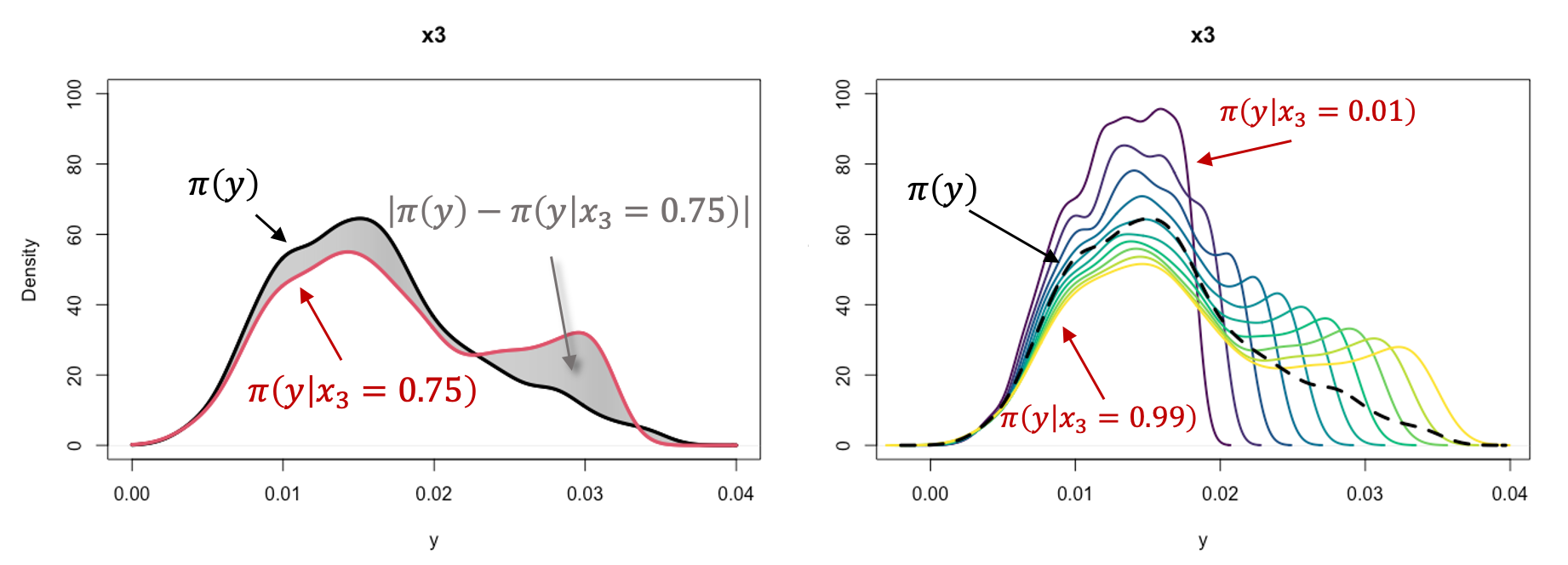}
         \caption{Intuition behind the delta index. In the left panel, the grey region is the change in PTW strength model response density (the inner integral) when $x_3$ is fixed at $0.75$.  The right panel shows how the density changes as $x_3$ is fixed at different values.  If a grey area were created from each line in the right panel to the dotted line, then $\delta_3$ would be estimated by the average of those grey areas (given the uniform prior on $x_3$).  }
         \label{fig:delta_intuition}
\end{figure}

Assume that $\bm x \sim \pi(\bm x)$ and consider the response a random variable $f(\bm x) = y \sim \pi(y)$. Define $\pi(y|x_i)$ to be the distribution of $f(\bm x)$ when the $i$th variable is fixed at $x_i$ but all other variables vary according to $\pi(\bm x)$.  Then the delta sensitivity is given by
\begin{align}
    \delta_i = \frac{1}{2}E_{x_i}\left[ \int_y |\pi(y) - \pi(y|x_i)| dy  \right] = \frac{1}{2} \int_{x_i} \pi(x_i) \int_y |\pi(y) - \pi(y|x_i)| dy dx_i,
\end{align}
which measures the change in the response density $\pi(y)$ when the $i$th variable is fixed at $x_i$ and the expectation is taken over the distribution of $x_i$. Figure \ref{fig:delta_intuition} shows the intuition behind this metric using $x_3$ from PTW. The left plot shows the difference between $\pi(y)$ and $\pi(y|x_3=0.75)$, indicating that the density when $x_3$ is fixed becomes bimodal. The area of the grey region is $\int_y |\pi(y) - \pi(y|x_3)| dy$ when $x_3=0.75$, and the right plot shows how $\pi(y|x_3)$ changes as $x_3$ is fixed at different values across its range. $\delta_3$ is then the expectation over the deviations between the densities when $x_3$ is fixed at different values.

The delta indices have the property that $0\leq \delta_i \leq 1$, with $\delta_i=0$ only when $f(\bm x)$ is independent of $x_i$ and $\delta_{1,\dots,p} = 1$. Different metrics for distance between densities can be used \citep{borgonovo2016sensitivity} to replace the inner integral. One of the most powerful properties of this metric is that it is invariant under monotone transformations of the function response. For instance, taking the log transformation of the function output will not change the delta indices, but substantially changes the interpretation of Sobol' indices.

\subparagraph{Application to the PTW Strength Model}

\begin{figure}[htbp]
         \centering
         \includegraphics[width=.6\textwidth]{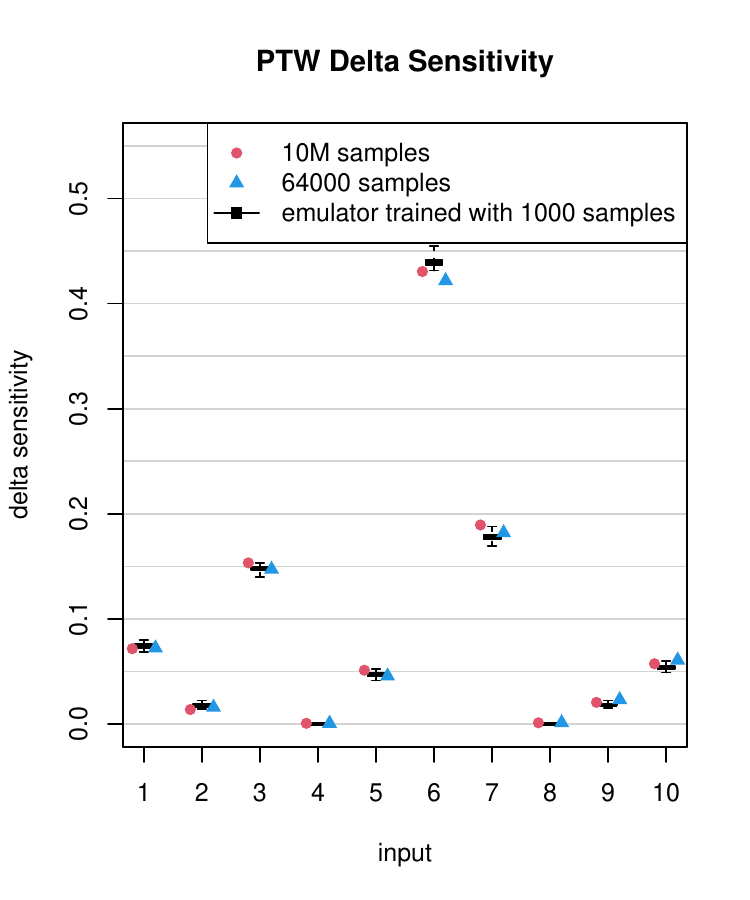}
         \caption{Delta indices under three different estimation scenarios.}
         \label{fig:delta_analysis}
\end{figure}

\begin{figure}[htbp]
         \centering
         \includegraphics[width=\textwidth]{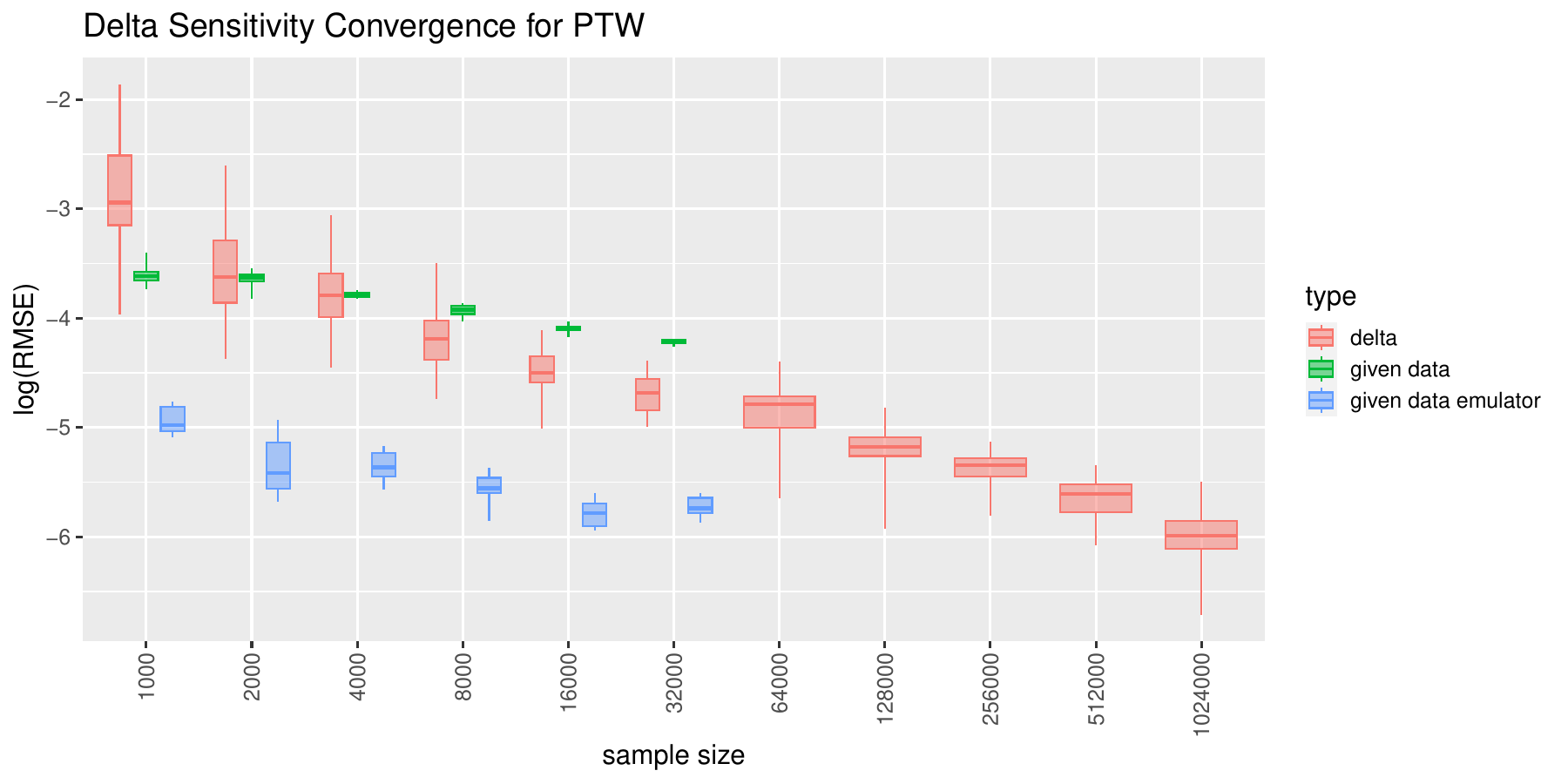}
         \caption{Convergence of delta indices as sample size increases. Emulation provides significant gains over the ``given data'' approach without an emulator.}
         \label{fig:delta_convergence}
\end{figure}

Figure \ref{fig:delta_analysis} shows the delta indices for PTW under three calculation scenarios: using ten million model runs for Monte Carlo calculation of the integrals (likely to be the ``truth'' in this case), using 64000 model runs, and using an emulator trained using 1000 model runs. There is decent agreement, though the emulator approach could be nearer to the truth than the approach with 64000 samples. Functions exist for calculating these indices when the model runs are fixed (i.e., the practitioner is given the data and cannot design the model runs to optimize sensitivity calculations) \citep{plischke2013global}. Figure \ref{fig:delta_convergence} shows how increasing sample size improves the accuracy of the estimated delta indices, and also shows that delta indices can be calculated very accurately from emulators. For the PTW strength model, using the BASS emulator results in roughly an order of magnitude more accuracy than using ``given data'' techniques (via python package \texttt{SALib} \citep{Herman2017,Iwanaga2022}).  Results were obtained using the indices calculated from a sample of size 10 million as the ``truth'' and taking the root mean squared error (RMSE) of the delta indices under 30 repeated samples for each sample size. 

Figure \ref{fig:shap_sob_del_dgsm} shows the delta indices for PTW compared to other single number summaries. There are substantial differences from the variance-based Sobol' and Shapley indices. Inputs 4 and 8 are still deemed unimportant, but inputs 6, 3, and 7 are more similar in their importance. Also, the importance ranking changes, with input 7 being the second most important rather than input 3.

\paragraph{Accumulated Local Effects (ALE)}

The ALE method \citep{apley2020visualizing} is especially relevant to the visualization of effects when inputs are dependent. Interaction effects can also be obtained with ALE.  To our knowledge, there is not yet an established sensitivity metric based on ALE (though variance of ALE effects is used in \cite{nakhleh2021exploring}), but ALE has become a popular choice for visualizing effects. 

\subparagraph{Mathematical Details}

The $i$th effect is constructed as follows:
\begin{align}
    ALE(x_i) = \int_{x_i^0}^{x_i} \int \frac{\partial f(\bm z)}{\partial z_i} \pi(\bm z_{-i}|z_i) dz_{-i}dz_i
\end{align}
which averages the local effect over all other variables (conditional on the location of the local effect) and then accumulates the averaged local effect.  We use $\bm z$ in place of $\bm x$ in the inner integral because $x_i$ is used in the bounds of the outer integral.  

\subparagraph{Application to the PTW Strength Model}

Some of the ALE effects for the PTW strength model, obtained using the R package \texttt{ALEPlot}, are shown in Figure \ref{fig:effects}. They look similar to the Sobol' effects plots in this setting, but are more appropriate when inputs are dependent.



\subsection{Not Enough Model Runs to Build An Emulator}

The sensitivity analysis approaches presented below are good choices when the modeler has limited resources that prevent sufficient model runs to build an emulator. Methods that require independent inputs are discussed first; methods that allow for input dependencies are covered second.

\subsubsection*{Requires Independent Inputs}

\paragraph{Morris Method}

The Morris Method \cite{morris1991factorial} is designed to help the modeler determine which inputs have important effects on the output of interest, given a moderate-to-large number of inputs, by considering \textit{elementary effects}, defined as the change in the output due solely to changes in a \textit{particular} input. 
For a given input $i$, if (1) the elementary effect is zero regardless of the values assigned to the remaining inputs, then input $i$ has a \textit{negligible effect} on the output; (2) the elementary effect is a nonzero constant across all values of the remaining inputs, then the effect of input $i$ on the output is \textit{additive and linear}; (3) the elementary effect is a nonconstant function of at least one of the inputs, then input $i$ is involved in higher-order effects and/or at least one interaction; the Morris Method does not distinguish between these.

The Morris approach is designed to enable the modeler to recognize whether each input belongs to category (1), (2), or (3) by considering the finite distribution of elementary effects. For input $i$, denote the finite distribution of elementary effects by $F_i$. A large absolute measure of central tendency for $F_i$ indicates that input $i$ has a large overall influence on the output, while a large measure of spread indicates that the degree to which input $i$ influences the output is dependent upon the values of other inputs, suggesting the presence of interactions or higher-order effects. 

Morris proposes using individually randomized one-factor-at-a-time sampling schemes to obtain a random sample of observed elementary effects. From these samples, estimates of the mean and standard deviation of $F_i$ are obtained and used to assess the magnitude of central tendency and spread for $F_i$. The Morris Method specifies a sampling plan with size that scales linearly with the number of inputs, thus requiring a relatively small number of runs.

\subparagraph{Mathematical Details}

The \textit{elementary effect} of input $i$ is defined as:

$$d_i(\mathbf{x}) = [y(x_1, x_2, \dots, x_{i-1}, x_i + \Delta, x_{i+1}, \dots, x_p) - y(\mathbf{x})]/\Delta$$

\noindent where $\mathbf{x} \in [0, 1/(k-1), 2/(k-1), \dots, 1]^p$, the $p$-dimensional $k$-level grid for predetermined $k$, $x_i \leq 1-\Delta$, and $\Delta$ is a predetermined multiple of $1/(k-1)$. For input $i$, denote the finite distribution $k^{p-1}[k-\Delta(k-1)]$ elementary effects by $F_i$. Morris proposes a sampling scheme to obtain samples from $F_i$ in an efficient manner as follows: 
\begin{enumerate}

    \item Construct a sampling matrix $\mathbf{B}$, an $m\times p$ matrix where all elements are 0 or 1 and for every column there are two rows for which only the $i$th column differs. 
    
    \item Obtain a randomized version of $\mathbf{B}$, denoted by $\mathbf{B}^*$, and defined as follows:

$$\mathbf{B}^* = \left(\mathbf{J}_{m\times1}\mathbf{x}^* + (\Delta/2)\left[(2\mathbf{B}-\mathbf{J}_{m\times p})\mathbf{D}^* + \mathbf{J}_{m\times p}\right]\right)\mathbf{P}^*$$

\noindent where $\mathbf{J_{m \times p}}$ is a $m \times p$ matrix of ones and $m \geq p+1$; $\mathbf{x}^*$ is a random starting value for $\mathbf{x}$; $\mathbf{D}^*$ is a $p$-dimensional diagonal matrix for which each element is $1$ or $-1$, assigned uniformly at random; and $\mathbf{P}^*$ is a random permutation matrix. Then $\mathbf{B}^*$ is a random orientation of $\mathbf{B}$, providing one randomly-selected elementary effect per input, where each of the $k^p/2$ elementary effects for input $i$ has equal probability of selection. 

\item Obtain $r$ independent random orientations of $\mathbf{B}$ to form the design matrix used by the modeler:

$$\mathbf{X} = [\mathbf{B}^*_1; \mathbf{B}^*_2; \dots; \mathbf{B}^*_r]^\textrm{T}$$

\end{enumerate}
$\mathbf{X}$ requires a total of $rm$ model runs. Extensions to the method exist that provide more than one elementary effect per input, thus requiring fewer model runs per effect.

Let $\bar{d}_i$ denote the sample mean and let $S^2_i$ denote the sample variance of the observed elementary effects for input $i$. Then $\bar{d}_i$ and $S^2_i$ serve as unbiased estimators of the mean and variance of $F_i$, respectively, and $\textrm{SEM}_i = S_i/\sqrt{r}$ provides an estimate of the standard error of the mean. To assess the relative importance of the inputs, Morris suggests constructing a plot that contains sample mean versus sample standard deviation for each sample of elementary effects. The experimenter can conclude that inputs obtaining high values of the sample mean have large main effects on the response, while inputs that obtain high values of both the sample mean and sample variance of the observed elementary effects are involved in a low-order interaction that impacts the response.

\begin{figure}[htbp]
         \centering
         \includegraphics[width=.6\textwidth]{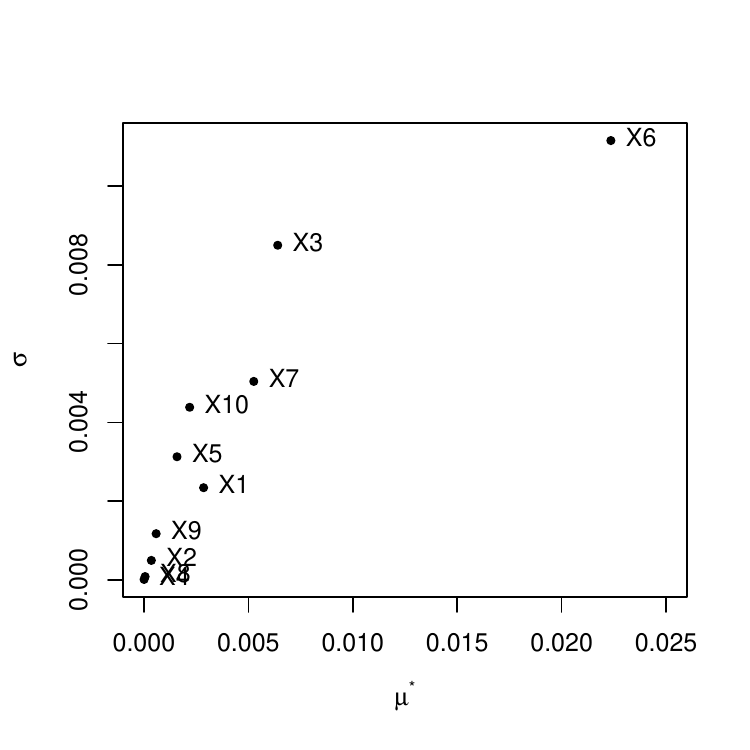}
         \caption{Application of the Morris method for PTW using 44 model evaluations.}
         \label{fig:morris}
\end{figure}

\subparagraph{Application to the PTW Strength Model}
Figure \ref{fig:morris} shows the results from applying the Morris method to PTW using the \texttt{morris} function from the R package \texttt{sensitivity}.

From these results, we note that $X_6$ obtains high values of both the sample mean (on the horizontal axis) and sample variance (on the vertical axis) of the observed elementary effects. Thus, we conclude that $X_6$ is involved in a low-order interaction that effects the response. Likewise, $X_3$ obtains relatively high values of sample mean and sample variance, providing some evidence that $X_3$ may also be involved in an important low-order interaction. Finally, $X_7$ obtains a somewhat high sample mean, suggesting $X_7$ also plays a role.

\paragraph{Derivative-Based Global Sensitivity Measure (DGSM)}
This metric is designed to provide bounds on the Sobol' total indices using few model evaluations. Various special cases have been pursued, including subset approaches.  Like almost all of the approaches we have discussed thus far, these account for interactions but do not detail them, though extensions have been pursued in \cite{roustant2014crossed}. We also note that active subspace based sensitivity connects to the DGSM approach \citep{constantine2017global}.

\subparagraph{Mathematical Details}

The DGSM  approach \citep{kucherenko2017derivative,lamboni2013derivative} is similar to the Morris method \cite{morris1991factorial} in that it will evaluate the gradient at many points in the input space and average, $w_i = E_{\bm x} \left[  \frac{\partial f(\bm x)}{\partial x_i} \right]$, or average the squared derivatives, $v_i = E_{\bm x} \left[ \left( \frac{\partial f(\bm x)}{\partial x_i} \right)^2\right]$.  Extensive work has been done to use DGSMs to bound Sobol' total indices especially under uniform and Gaussian input distributions, which can be powerful because DGSMs require significantly fewer model runs to estimate accurately.  

ALE is similar to the DGSM $w_i$ when inputs are independent, but instead of averaging over the $i$th dimension, it accumulates, because the main interest is visualization.  

\subparagraph{Application to the PTW Strength Model}

We examine the upper bounds for the Sobol' total indices in Figure \ref{fig:shap_sob_del_dgsm}, obtained using the \texttt{dgsm} function in the python library \texttt{SALib} with 200 model runs. The upper bound for input 6 extends up to 1.6, making it less of a useful bound. However, bounds for inputs 4 and 8 correctly indicate that those inputs do not contribute, and bounds for the rest of the inputs indicate that there are not large effects to compete with the effect of input 6.  We note that, because of how quickly averages like $w_i$ and $v_i$ can converge, emulators often perform poorly by comparison, though \cite{de2016estimation} find it useful to use Gaussian processes in some scenarios.  Tree-based emulators will perform especially poorly because of their discontinuous nature (and hence poor gradient estimation).  However, when an accurate emulator is available, we are unlikely to need DGSMs because we can estimate Sobol' total indices directly.

\paragraph{Design of Experiments-Based Approaches}

Design of Experiments-based approaches can be used to help the modeler identify low-order effects on the output of interest, given a relatively small number of available model runs. A common objective in early stages of experimentation is to identify the inputs that most affect the response of interest. Often, the experimenter starts with a relatively large collection of potentially-important inputs and seeks to select those that are most important, using a limited number of model runs. Screening designs, which encompass a class of Design of Experiments-based approaches, have traditionally been used for this purpose in industrial and other physical experiments. Morris \cite{morris1991factorial} notes that computational experiments can likewise benefit from the application of screening design methodologies when a limited number of model runs is available.

Data collected via a screening design is fit using an approximating low-order linear model. Effects deemed most important are those with large attendant model coefficients. Inputs associated with identified important effects are retained for future phases of study. This approach as been widely and successfully used for decades to study physical systems across myriad domains \cite{montgomery}. Morris \cite{morris1991factorial} observes that, although computational experiments lack physical random quantities, this approach has nevertheless been shown to work well in practice for computational experiments.

Traditionally, experimenters relied on screening designs such as resolution III or IV fractional factorial designs \cite{BoxAndHunter} or Plackett and Burman designs \cite{PlackettBurman}. However, these designs do not enable the experimenter to detect distinct two-factor interactions or to identify the presence of curvature \cite{montgomery}. As Morris \cite{morris1991factorial} notes, this presents potentially serious disadvantages, particularly in the context of computational experiments.

Jones and Nachtsheim \cite{DSDs} introduced a class of screening designs called Definitive Screening Designs (DSDs) that avoid these disadvantages. In this section, we provide a brief introduction to these designs and investigate their performance for sensitivity analysis via the PTW Strength Model. Additional Design of Experiments-based methods like Bayesian D-optimal supersaturated designs \cite{supersaturatedDopt} can also be useful for sensitivity analysis with limited model runs available, and we encourage the reader to consider a variety of Design of Experiments-based approaches and identify one best-suited for the needs of the specific problem at hand. In our experience, the advantages of DSDs make them a nice starting point, and so we focus on these designs in this overview.

\subparagraph{Definitive Screening Designs}

DSDs \cite{DSDs} are a class of three-level screening designs requiring a relatively small number of runs: for $p$ continuous inputs, DSDs require a minimum of only $2p + 1$ runs. These designs consist of $p$ fold-over pairs and one overall center run, leading to three center values per column. The structure of these designs results in several key advantages: 
\begin{enumerate}
    \item A low number of runs is required (only one more than twice the number of inputs)
    \item Unlike resolution III fractional factorial designs or Plackett and Burman designs, main effects are completely independent of two-factor interactions, which enables main effects to be estimated separately from two-factor interactions
    \item Unlike resolution IV fractional factorial designs, two-factor interactions will never be completely confounded with other two-factor interactions, which enables the experimenter to detect distinct interaction effects
    \item Unlike traditional two-level screening designs, all pure quadratic terms are estimable, enabling the detection of curvature and the ability, potentially, to screen and optimize in a single step. 
\end{enumerate}

\subparagraph{Mathematical Details}

While originally constructed algorithmically, Xiao et al. \cite{Xiao2012} showed that, for many even values of $p$, a minimum-run, $p$-factor DSD, $\textbf{D}_{p, \textrm{min}}$, can be constructed by taking

$$\textbf{D}_{p, \textrm{min}} = \begin{pmatrix}
    \textrm{C}_p \\
    - \textrm{C}_p \\
    \mathbf{0}
\end{pmatrix}$$
\noindent where $\textbf{C}_p$ is a $p \times p$ conference matrix and $\mathbf{0}$ is a row vector of zeros. A conference matrix is an orthogonal matrix such that $\textbf{C}_p^\prime\textbf{C}_p = (p -1)\textbf{I}_p$. For odd values of $p$, the conference matrix approach should be used to obtain a design for $p^\prime = p + 1$ factors. The extra column is simply dropped from the design to obtain a $p$-factor DSD consisting of $2p +3$ runs. Using this approach guarantees orthogonality of the main effects.

Jones and Nachtsheim \cite{DSDsWithCatFactors} presented two methods for augementing DSDs for use when the experimental inputs consist of a mixture of two-level categorical factors and continuous factors. Nachtsheim et al. \cite{twoLevelAugmentedDSDs} proposed methodology for obtaining nondominated, Pareto-optimal designs from the class of two-level augmented DSDs.

DSDs are supersaturated designs with respect to the full second-order linear model. In their investigation of model selection methods for supersaturated designs, Marley and Woods \cite{MarleyAndWoods} suggested the use of Gauss-Dantzig selector and recommended that the number of runs be three times the number of active factors. In their comparison of model selection methods for DSDs, Errore et al. \cite{errore2017} concluded similarly that Lasso and Gauss-Dantzig performed best and noted a marked drop off in performance across all model selection methods when the number of active effects was more than about half the number of runs in the design. To mitigate this risk, Errore et al. recommended the use of two fake factors, yielding a $p^{\prime\prime} = p + 2$-factor design in $2p + 5$ runs \cite{errore2017}. The columns corresponding to the fake factors can then be dropped from the design to obtain the desired $p$-factor design.

Jones and Nachtsheim \cite{fittingDSDs} demonstrated a model selection method that exploits the  structure of DSDs to outperform standard model selection methods. Leveraging work by Miller and Sitter, \cite{millerAndSitter} Jones and Nachtsheim recommended an approach that separates the analysis of the main effects (odd functions) from the analysis of two-factor interactions and pure quadratic effects (even functions). A function $f$ is (1) an odd function if $f(-x) = f(x)$ for all $x$ or (2) an even function if $f(-x) = f(x)$ for all $x$. Because DSDs are foldover designs, it is possible to separate the sample space for the response into (1) the sample space spanned by the odd effects and its orthogonal compliment (2) the space spanned by the even effects. As Jones and Nachtsheim show, the use of this approach results in considerable gains in the power to detect second-order effects and is the recommended approach to model fitting for DSDs.

\subparagraph{Application to the PTW Strength Model}

For the PTW strength model, there are ten parameters of interest. Following the recommendation of Errore et al. \cite{errore2017} we constructed a DSD for ten factors and two fake factors, yielding a $2p + 5 = 25$-run design. These 25 model runs were executed and the corresponding response values obtained. We employed the design-based model selection method proposed by Jones and Nachtsheim \cite{fittingDSDs}. The procedure obtains the following active effects: $X_1, X_3, X_5, X_6, X_7, X_1X_6, X_3X_6, X_5X_6, X_6X_7$, and $X_6^2$. The sorted parameter estimates are given in Table \ref{tab:DSD_parameter_estimates}.

\begin{table}[]
    \centering
    \begin{tabular}{ccccc}
Term & Estimate & Std Error & $t$ Ratio & Prob $> |t|$ \\
 \hline
$X_6$ & -0.0089 & 0.00032 & -27.57 & $<.0001$ \\
$X_3$ & 0.0042 & 0.00032 & 12.91 & $<.0001$ \\
$X_3X_6$ & -0.0048 & 0.00038 & -12.21 & $<.0001$\\
$X_7$ & 0.0028 & 0.00032 & 8.68	& $<.0001$ \\
$X_6X_7$ & 0.0029 & 0.00035 & 8.47 & $<.0001$ \\
$X_5$ & -0.0011 & 0.00032 & -3.39 & 0.0049 \\
$X_1$ & 0.0011 & 0.00032 & 3.36 & 0.0051 \\
$X_1X_6$ & 0.0011 & 0.000351 & 3.23 & 0.0066 \\
$X_5X_6$ & -0.0010 & 0.00035 & -2.93 & 0.0117 \\
$X_6^2$ & 0.0029 & 0.00098 & 2.91 & 0.0122 \\ 
$X_5X_7$ & -0.0011 & 0.00039 & -2.72 & 0.0175 \\
    \end{tabular}
    \caption{Sorted parameter estimates for 25-run DSD-based screening study of PTW strength model}
    \label{tab:DSD_parameter_estimates}
\end{table}

From Table \ref{tab:DSD_parameter_estimates}, we note that including $X_6$ in the model has the largest effect on accounting for the variability in the response. In fact, we find that the model that contains only $X_6$ explains 58\% of the variability in the response. $X_3$ and the $X_3X_6$ interaction account for the second and third most important effects. The model including $X_6$, $X_3$ and $X_3X_6$ accounts for 85\% of the variability in $Y$. Adding $X_7$ and the $X_6X_7$ interaction brings the total variability explained by the model to 95\%: clearly, $X_3$, $X_6$, and $X_7$ are critical inputs in the PTW model. Including $X_5$, $X_1$, the $X_1X_6$, $X_5X_6$, and $X_5X_7$ two-factor interactions, and the pure quadratic effect in $X_6$ obtains a model that explains over 99\% of the variability in the response.

Employing the DSD Design of Experiments-based approach, enabled identification of the most important inputs, including the ability to characterize two-factor interactions and curvature in the form of pure quadratic effects, all using a highly limited number of model runs: this approach required only 25 model runs to study ten inputs. This is far below the numbers required by other methods covered here, yet, in this case, the results obtained agree with more resource-intensive approaches and, compared to some approaches, they provide additional information in their ability to estimate interaction and pure quadratic effects. 

\subsubsection*{Allows Dependent Inputs}

\paragraph{Constrained Optimal Design-Based Approaches}

Design of Experiments-based approaches can be applied in cases where the inputs are constrained by some dependencies. In these cases, optimal design-based approaches provide the flexibility required to address constraints induced by dependent inputs. 

Optimal design of experiments is a flexible method for designing experiments that allows the design to be tailored to the system under study. Optimal designs are constructed to maximize a criterion of interest, defined to be a function of the variance-covariance matrix of the ordinary least squares estimator \cite{fedorov}. The criterion of interest is selected to best support the needs of the experiment; some criteria are most appropriate when the experimenter is primarily interested in parameter estimation and others are best for when the experimenter's objective is to obtain precise predictions of a response of interest. In either case, optimal design of experiments methodology supports additional tailoring of the design, including the ability to impose linear constraints in the inputs, to account for dependencies.

For instance, it may be the case that a collection of inputs should not simultaneously be set to their lowest or highest setting. This configuration may be infeasible or it may simply be unrealistic and therefore not of interest to study. In such cases, inequality constraints can be imposed on the inputs to avoid combinations that should be excluded \cite{atkinson2007}. Optimal design of experiments approaches can be employed to determine the input combinations at which the modeler should exercise the system to identify low-order effects while avoiding infeasible or uninteresting settings.

Similarly, optimal design of experiments approaches can accommodate scenarios in which some or all inputs must sum to a constant. These so-called mixture experiments exhibit dependencies in the inputs that can be managed using optimal design of experiments approaches. Mixture experiments necessitate additional considerations at the analysis stage. A model that is commonly used for the analysis of data from mixture experiments is the Scheffé model. If only a subset of the inputs are constrained to sum to a constant, then a low-order polynomial model can be crossed with a Scheffé model to account for the presence of both dependent and independent inputs \cite{millerAndSitter}.

\section{Discussion}

\begin{figure}[htbp]
         \centering
         \includegraphics[width=\textwidth]{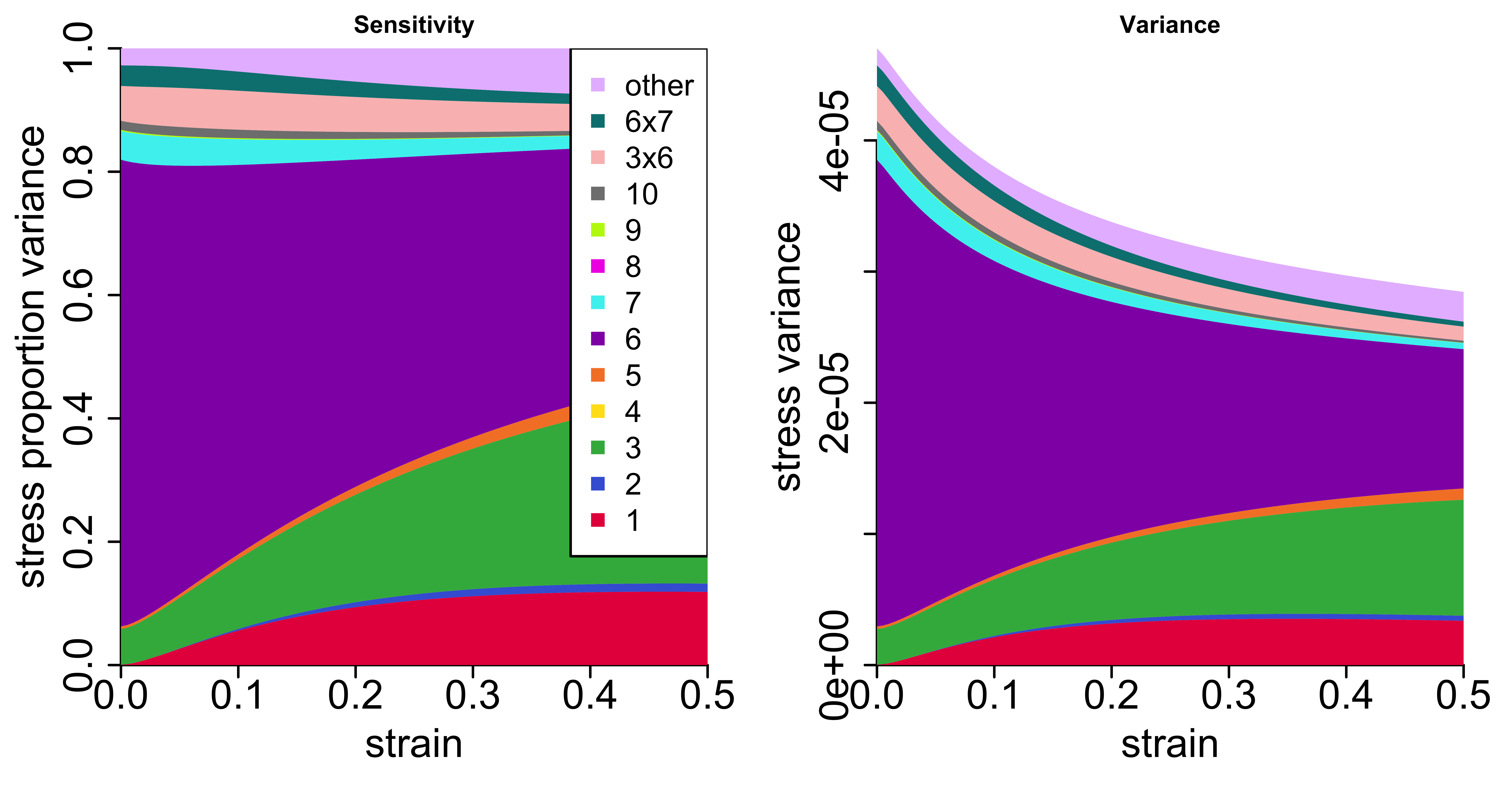}
         \caption{Functional Sobol' decomposition using the \texttt{sobolBasis} function from R package \texttt{BASS}. The left plot shows the main and interaction effects of the typical Sobol' decomposition in a functional pie chart, while the right shows the same effects scaled by the function variance (the variance of the functional responses decreases as strain increases).}
         \label{fig:sobol_functional}
\end{figure}

We treated the PTW strength model in a simplified way so that all analyses could be compared: we made inputs independent and we disregarded the functional response.
To do sensitivity analysis for multivariate or function response, most methods will just repeat the analysis for each response.  If the number of responses is large, this can lead to complications for computation or visualization intensive approaches.  \cite{lamboni2011multivariate} use a PCA-based approach to reducing the dimension, while \cite{francom2019inferring} use a PCA-based approach to build an emulator and perform sensitivity analysis using the emulator.  This same approach is used to make the ``functional pie charts'' in Figure \ref{fig:sobol_functional} for PTW.  The analyses in the section above were performed at strain of 0.05, and the Sobol decomposition in Figure \ref{fig:sobol_functional} shows how the sensitivity changes as a function of strain.

We also took for granted that, in global sensitivity analysis, we must choose an input distribution.  Recent work has tried to limit sensitivity of the sensitivity analysis to the choice of distribution in a rigorous way \citep{borgonovo2018functional,schobi2019global} allowing for multiple input distributions to be considered, and this is an active area of research.  This is again an area where emulators are helpful, since a single emulator could be fit in such a way that multiple sensitivity analyses (under different input distribution assumptions) could be performed.

In this chapter, we provided an overview of widely-used sensitivity analysis methods that we hope will benefit the practitioner. We considered a variety of global sensitivity methods that provide flexibility to the modeler, address a range of practical considerations, and are widely available in standard software packages. We hope that by providing guidance on sensitivity analysis approaches framed in the context of a real example -- the PTW strength model -- and by incorporating the critical issues that guide the selection of a sensitivity analysis method -- available resources, input dependencies, and desired information -- that this chapter may serve as a useful resource and a helpful decision-making tool.

\bibliographystyle{plain}
\bibliography{bibl}

\end{document}